\documentclass[printer]{aa}
\usepackage{graphics,epsf,epsfig,graphicx}
\begin{document}

\title{ Formation and evolution of galactic disks with a multiphase numerical model}

\author{B. Semelin \inst{1} \and  F. Combes \inst{2}} 
\offprints{B. Semelin, semelin@gravity.phys.waseda.ac.jp} 
\institute{
Department of Physics, Waseda University, Oh-kubo, Shinjuku-ku, Tokyo 169-8555, Japan
\and
Observatoire de Paris, DEMIRM, 61 Av. de l'Observatoire, 
F-75014, Paris, France 
} 
\date{Received XX XX, 2002; accepted XX XX, 2002} 
\authorrunning{Semelin \& Combes} 
\titlerunning{}

\maketitle

\abstract{ The formation and evolution of galactic disks are complex
phenomena, where gas and star dynamics are coupled through 
star formation and the related feedback. The physical processes are so
numerous and intricate that numerical models focus, in general, on
one or a few of them only. We propose here a numerical model with particular
attention to the multiphase nature of the interstellar medium; we consider a
warm gas phase ($\ge$ 10$^4$K), treated as a continuous fluid by an
SPH algorithm, and a cold gas phase (down to 10K), fragmented in clouds,
treated by a low-dissipation sticky particles component.
The two gas phases do not have the same dynamics, nor the same spatial
distribution. In addition to gravity, they are coupled through mass exchanges
due to heating/cooling processes, and supernovae feedback. Stars
form out of the cold phase, and re-inject mass to the warm phase through
SN explosions and stellar winds. The baryons are embedded in a live
cold dark matter component. \\
Baryonic disks, initially composed of pure gas, encounter
violent instabilities, and a rapid phase of star formation, that slows
down exponentially. Stars form in big clumps, that accumulate in the
 center to build a bulge. Exponential metallicity gradients
 are obtained. External infall of gas should be included to maintain
 a star formation rate in the disk comparable to what is observed
 in present disk galaxies.
 }

\section{Introduction} 

The formation of a galactic disk involves such complex physical
processes, that only numerical computations can give some insight
in their relative importance. The essential force is gravity, which
is nowadays easily controlled through N-body codes, based either
on grid calculations (such as FFT), or on hierarchical grouping of
particles (such as the tree-code). Both algorithms optimize
the computing time, growing roughly as $N \ln(N)$. The main improvement in this
domain is obtained by going towards higher and higher spatial resolution,
with an ever growing number of particles. The gravitational force is
softened at small scale, to reduce unphysical two-body
relaxation, resulting in a resolution equal or slightly
smaller than the inter-particle distance (Romeo \cite{Romeo98},
Dehnen \cite{Dehnen}, Knebe at al. \cite{Knebe01}).

It has been recognized for a long time that the dissipative component
is also an essential feature in galaxy evolution, and the
interstellar medium has been widely introduced in galaxy
simulations, either as a continuous fluid (Eulerian grid codes,
van Albada \& Roberts  \cite{vanAlbada81},
Lagrangian SPH codes, Hernquist and Katz  \cite{Hernquist89}), or through
sticky particles algorithms (Combes \& Gerin \cite{Combes85}). However, contrary
to the pure gravity case, the result of simulations now depends not only on
the spatial resolution and computer power, but on the physical
assumptions made on the nature of the ISM and related processes.

Star formation and feedback, coupling the stars and gas by non-gravitational
processes are essential (Katz  \cite{Katz92},
Mihos \& Hernquist \cite{Mihos}). Since the detailed processes
involved in star formation are not yet well known, this introduces
further uncertainties and liberties in the modeling. The most widely adopted
recipe to control star formation is based on a local Schmidt law (i.e.
the star formation rate is locally proportional to some power of the volume
density), sometimes associated with a threshold for star formation
(e.g. Friedli \& Benz \cite{Friedli95}).
However, the local Schmidt law is not actually observed in galaxies,
and the justification comes from an observed global empirical Schmidt law,
averaged over the whole galaxy (e.g. Kennicutt \cite{Kennicutt98}).
  Other star formation recipes are based on Jeans instability
  (e.g. Steinmetz \& M\"uller \cite{Steinmetz}, Gerritsen \& Icke \cite{Gerritsen}),
  or cloud-cloud collisions (Noguchi \& Ishibashi \cite{Noguchi86}).

Due to the limited number of particles, each star ``particle'' represents in fact
a stellar cluster of the order of 10$^5$ to 10$^6$ M$_\odot$.  The conversion
of a fraction of a gas particle into stars requires the creation of a large
number of star-particles (e.g. Katz \cite{Katz92}), or the consideration of
``hybrid'' particles, that are transiently containing both gas and stars,
and have the same dynamics for a while (Mihos \& Hernquist \cite{Mihos}).
Alternatively, some ``starlet'' particles are transiently created, and merged
with the nearest neighbors (Jungwiert et al. \cite{Jungwiert}). Whatever
the choice, some approximations are made, related to the limited resolution
in time, mass and space, respectively of the order of
10$^6$ yr,  10$^6$ M$_\odot$ and 0.3 kpc for a typical giant spiral galaxy.

The same limitations occur when mass loss and feedback are considered.
Either only stellar heating is considered (Gerritsen \& Icke \cite{Gerritsen}),
or the supernovae mechanical energy only(Mihos \& Hernquist \cite{Mihos}),
or more sophisticated models are used including several types of supernovae, 
planetary nebulae, stellar winds, evaporation and condensation
(e.g. Theis et al. \cite{Theis92}, Samland et al. \cite{Samland97},
Berczik \cite{Berczik99}).

Stellar nucleosynthesis can be followed to describe the detailed metal enrichment
history of the galaxy (e.g. Lia et al. \cite{Lia}). Models reveal that the
frequently used instantaneous recycling is a too simple approximation
(see also  Jungwiert et al. \cite{Jungwiert}).

Of first importance is the thermal evolution of the gas, since
thermal instabilities are responsible for cloud condensation,
feeding the cold gas phase, and thus star formation (Hultman \&
Pharasyn \cite{Hultman}). Cooling
and heating processes are taken into account depending on the temperature,
density and metallicity of the gas (e.g. Dalgarno \& McCray \cite{Dalgarno72},
Sutherland \& Dopita \cite{Sutherland}). However, the time-scales
involved in the thermal processes can be much smaller than the dynamical
time-scales, and the spatial scales of the corresponding processes
are much below the spatial resolution of the simulations. Phenomenological 
recipes are therefore used to convey at large-scales the resulting effects of
the processes occurring below the resolution (e.g. Thacker \& Couchman
\cite{Thacker00}): the heating energy is smoothed over the resolution
scale, and the effective time-scales slowed down.

Many simulation models consider only one gas phase, treated as
a continuous fluid at the virial temperature corresponding
to a galaxy potential ($\sim$ 10$^4$ K). A simple approximation
is a strictly isothermal gas, since the cooling is very efficient above
10$^4$ K (Barnes \& Hernquist \cite{Barnes91}, Mihos \& Hernquist
\cite{Mihos}). Some allow gas to cool down to low temperatures
($\sim$ 10 K), and to spread  into several temperature phases
(e.g.  Gerritsen \& Icke \cite{Gerritsen}; Yepes et al. \cite{Yepes},
Thacker \& Couchman \cite{Thacker01}). However, their dynamics
is still that of a one-phase medium, given the insufficient density
contrast that can be dealt with the numerical simulations on large-scales.
Significant contrasts can be achieved only in simulations of small volumes,
that approach more realistically the multiphase nature of the
interstellar medium (e.g. Rosen \& Bregman (\cite{Rosen}),
Wada \& Norman (\cite{Wada99}, \cite{Wada01}).

In the present paper, we implement most of the above processes
to investigate the formation and evolution of disk galaxies
(gravity, stellar and gas dynamics, star formation, feedback and massloss
heating and cooling, metal enrichment..), and we focus on the
large-scale multi-phase dynamics of the gas. The ``warm gas''
component is modeled by a continuous fluid with an SPH
code, and contains a large range of temperatures, up to hot gas
re-injected by supernovae, though the bulk of the gas is found
between 10$^4$ and 2 10$^4$K. The ``cold gas'' component
corresponds to the cloudy and fragmented, essentially molecular
medium, between 10K and 100K. It is modeled as a separate phase,
via individual cloud-particles, subject to inelastic collisions. Since at large
galactic scales, the density contrast necessary to form such a
fragmented structure (12 orders of magnitude)
is far beyond the present computational
power, the cloud-particle approximation appears well suited. For
the cloudy component, the pressure forces or viscosity forces
are negligible.  The dissipation occurring during cloud collisions
at AU scales, much below our spatial resolution, is modeled
phenomenologically.

Two general classes of models have been proposed for the
formation of galaxies: either big galaxies can form through
the monolithic collapse of a very massive baryonic system
(e.g. Eggen, Lynden-Bell \& Sandage 1962), or
dwarf galaxies form first, and act as the building blocks
of larger systems, that subsequently form through recursive mergers,
in the hierarchical scenario (e.g. Kauffmann et al 1999).
The first scenario assumes that the gas experiences violent 3D star
formation, so quickly that it has no time to settle into a disk.
This has been proposed to form large elliptical galaxies or bulges.
However, it is today recognized that many (if not all) elliptical
galaxies can form through mergers of spiral galaxies, or smaller
entities,
as first suggested by Toomre \& Toomre (1972), and developed further
by Schweizer (1986), while bulges can form similarly by mergers,
and also through secular evolution (e.g. Combes 2000).
Although the most successful cosmological scheme today, based on
inflation,
promotes primordial density fluctuations that are scale-invariant, which
together with the existence of cold dark matter, results in hierarchical
formation of structures, the collapse of baryons inside dark matter
halos is still an unsolved problem, with many uncertainties and free
parameters. The rate at which stars are formed in a given system,
and whether the stellar populations observed today can be attributed to
a sudden event like the monolithic collapse, or a more progressive one
like the gradual and recursive merging, is a matter of debate. It is
likely
that both point of views may correspond to some observed galaxy examples
today, since the major merger of two gas rich giant spirals have some
similarities with a monolithic event.  More subtle diagnoses should be
searched for, and large number statistics should be used to determine
the dominant processes. In this work, we focus on the formation and evolution
of a single galaxy. To achieved the required resolution we use the monolithic
collapse scenario.

In section 2 we detailed the assumptions of the multiphase model, and
present the code and numerical methods in section 3. In section
4, 5 and 6 we present and discuss the results from our simulations.
We give our conclusions in section 7.

\section{The multiphase model}

We will describe in detail the different aspects of the model. Fig. 1
presents a schematic view for quick reference. 

\begin{figure}[t]
\resizebox{\hsize}{!}{\includegraphics{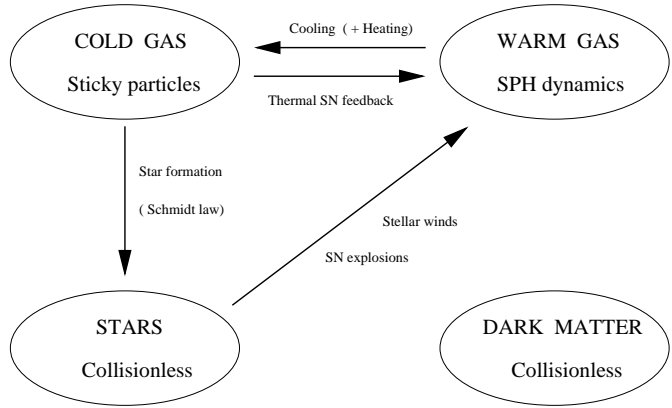}}
\caption{Schematic presentation of the model. The four phases are represented
with the main physical processes responsible for mass and energy exchanges.}
\end{figure}

\subsection{Description of the four components}

\subsubsection{Dark matter}
We have compared the results of two usual methods to take dark matter into 
account. The first uses an analytic, static potential. The second uses 
collisionless particles. The first choice obviously has a major shortcoming: the
dark matter density field does not respond to the evolution of the baryonic
matter. The static density configuration of the dark matter affects the 
evolution of the system,
and phenomena such as transfer of angular momentum from baryonic matter
to dark matter are ignored. While using a static potential may be
considered in the case we focus on in this work; the evolution of a single 
galaxy within a dark matter halo, it would appear
unreasonable in the case of mergers where dark matter halos undergo large 
deformations. We will show in this paper that some difficulties also appear
when using collisionless dark matter particles  with insufficient
numerical resolution. We show in Sec. 5 that using a ``small'' number
of heavy dark matter particles results in unphysical heating of the stellar
disk within 2 Gyr. This problem can be overcome partly by using large 
smoothing lengths for the dark matter, but more reliably by increasing the 
number of particles.

\subsubsection{Stars}
Stars form the second type of collisionless matter. They interact with
the gas components through processes described in Sec. 2.2 . In this work,
the mass of each stellar particle is of the order of $\sim 10^6 M_{\odot}$ 
in most
simulations and $\sim 10^5 M_{\odot}$ in the high resolution simulation. Each
particle thus represents a large number of stars, and such properties as
supernovae rate or metallic enrichment will depend on the assumed initial mass
function (see Lia et al. \cite{Lia}). In this work, we do not study
the detailed dependence on the IMF, but rather use standard values for 
quantities such as supernovae rate. Our model includes many processes and we 
choose a simple representation for each of them.

\subsubsection{Warm gas}
The warm gas in the ISM is a highly complex system. Dynamically, it can be
described in first approximation as a self-gravitating dissipative fluid.
In our model, the warm gas phase is composed of SPH particles obeying the 
ideal gas equation of state:

\begin{equation}
\qquad P=(\gamma -1) \rho u \qquad ( \gamma = {5 \over 3} ). 
\end{equation}

\noindent
Here, $P$ is the pressure, $\rho$ the density, and $u$ the thermal energy per
unit mass. The thermal energy is governed by equation:

\begin{equation}
\qquad \rho {du \over dt}+ \tilde{P} \nabla \hbox{\bf v}= 
n_H\Gamma_{\hbox{\tiny UV}}+ n_H^2 \Lambda(u), 
\end{equation}

\noindent
where $\hbox{\bf v}$ is the velocity of the gas and $n_H$ the number density
of hydrogen. $\tilde{P} $ is the viscosity-corrected pressure (see Sec.
3.1 for details and references on the SPH implementation).
On the right hand side are heating and cooling terms described in detail
in Sec. 2.2.1. The evolution of the internal energy is limited to the 
corresponding  finite range in temperature $11000 $K -$10^5$ K. 
The lower limit corresponds to the
transition to the cold star-forming gas phase. The upper limit
is for numerical convenience, since in this work most SPH particles lie in the 
11000 K - 20000 K range, at the virial temperature of typical galaxies. 

\subsubsection{Cold gas}

A number of numerical models stop the cooling of SPH particle
at 10000 K (e.g., Weil {\sl et al} \cite{Weil}, Thacker \& Couchman \cite{Thacker00}). This can only be a first approximation.

Indeed, a large amount of the gas in galaxies is observed in cold ( 10-100 K ) 
dense
molecular clouds. These objects are essential to the global dynamics of the 
galaxies since they are the location of star formation. Their physical 
properties
and dynamics differ strongly from those of the surrounding diffuse warm gas.
Indeed, each molecular cloud is in local virial equilibrium and the overall
filling factor of the cold gas is very small. Consequently, it cannot be
accurately described as a fluid (which supposes a rather continuous medium).
Moreover, current treatment of SPH uses the ideal gas state equation to relate
pressure, density and internal energy of particles which, 
at present days numerical resolution, have the mass of a
giant molecular cloud. This ideal gas
description is hardly adequate for cold, self-gravitating, GMC-like particles,
which should rather obey Larson laws (Larson \cite{Larson}). For these reasons, we feel
that a specific non-SPH treatment should be given to these star-forming gas 
particles.

Some attempts have been made in this direction. Let us first mention the works
by Wada \& Norman (\cite{Wada99}) and Wada (\cite{Wada01}) who include
proper cooling down to 10 K in Eulerian simulations of small regions of the
galaxy ( $\sim$ 100 pc). Yepes et al. (\cite{Yepes})
, using a PPM scheme, and Hultman \& Pharasyn (\cite{Hultman}), using a SPH 
simulation, both identify a phase of cold, star-forming gas. However,
both still treat the dynamics of the gas as a single fluid. Relying on the
fact that the largest fraction of the gas in the ISM is in the form of
cold clumpy fragments (molecular clouds), Noguchi (\cite{Noguchi}) adopts a 
sticky particle scheme to describe the collisional gas.  He does not account for
the warm diffuse gas. Andersen \& Burkert (\cite{Andersen}) develop a more
elaborate model to treat the cold gas clouds as a collisional fluid. They 
take warm gas into account only through a constant external pressure exerted on
cold clouds.
In this work we attempt to fully model a cold and a warm gas phase with
different dynamics.

While we describe the warm gas with the usual SPH technique, the cold gas
particles experience gravitational forces only, and are subject 
to inelastic collisions following the sticky particle scheme (Levinson \&
Roberts \cite{Levinson}). Within one time step, collisions occur with a 
probability 
$ P \sim \rho_{\hbox{\tiny loc}}\delta \hbox{v}$, where $\rho_{\hbox{\tiny loc}}$ is the local cold
gas density, and $\delta \hbox{v}$ is the local velocity dispersion. The only
"thermal" evolution for cold gas particles is the possibility to be
evaporated back to the warm gas phase through supernovae feedback (see Sec. 2.2.3)

\subsection{Physical processes}

\subsubsection{Cooling and heating}

The evolution of the thermal energy of warm gas particles follows equation (2).
The sources
of variation of the thermal energy are the following: adiabatic expansion, 
heating from viscosity, heating from an external radiation field (the $\Gamma_{\hbox{\tiny UV}}$
term) and radiative cooling  (the $\Lambda$ term).

We use the tabulated values of $\Lambda$ in Sutherland \& Dopita 
(\cite{Sutherland}). We have not
included the effect of metallicity on the cooling: we use a constant solar 
metallicity in the cooling function. Indeed, most of the gas lies in the 
10000K-20000K range, where metallicity has little effect. In the typical
conditions of galactic dynamics, the cooling time can become much shorter than 
the dynamical time in resolved dense regions. Reducing the dynamical time step 
accordingly is not appealing since it would drastically increase de CPU 
requirements. 
Moreover, it would not be consistent with the time scale resolution of 
gravitational effects. Consequently we choose to damp the thermal fluctuations
(the same solution is applied by Weil et al. \cite{Weil}). Whenever the 
variation of the internal energy of a particle within one time step is larger 
than one fourth of its current value, it is damped to one fourth. 
In addition, adaptative small steps are used 
to follow the very fast variations of 
$\Lambda$ as a function of $u$ in the 
10000K-20000K range  (this is similar to the integral scheme use by 
Thomas \& Couchman \cite{Thomas}). This results in a stable thermal behavior
for the gas.

A constant uniform heating $\Gamma_{\hbox{\tiny UV}} \sim 10^{-24}$ erg s$^{-1}$ is applied to 
the gas to model the background UV radiation field. The variations of
$\Gamma$ only shifts the mass balance between warm and cold phase as other
parameter can, such as the minimal temperature of the warm gas.
It is however
important to include a non-zero $\Gamma$ to help sustain the warm phase.
In our study we keep $\Gamma$ constant. Gerritsen \& Icke (\cite{Gerritsen})
propose a detailed model for the heating.

The second source of heating comes from supernovae feedback. This is the only
way in our model for cold gas to return directly to the warm gas state.
It will be described in Sec. 2.2.3. 

Let us explained now how the transition from warm gas to cold gas is handled.
A parameter $u_0$ is set which defines the thermal energy at the transition.
It is a free parameter, which value corresponds to a temperature of 
11000 K. A simple method would be to transform
into cold gas any warm gas particle cooling down to $u_0$. However, owing to
the different density dependence of cooling and heating terms in eq (2), this
choice induces a sharp density threshold for turning warm into cold gas.
When the average warm gas density falls noticeably below the threshold the
transfer to cold gas is stopped and the cold gas phase is progressively
depleted by star formation. We obtained better 
equilibrium between the phases with a smoother transition. Accordingly, and
for the purpose of transition to cold gas only, the thermal energy of each SPH 
particle of warm gas is considered as a Gaussian distribution around a
central value $u$ with a dispersion of $\sigma_u$, rather than a Dirac 
distribution $\delta(u)$.
This makes sense from a physical point of view since the
scale resolution of the simulation is about 100 pc, leaving out thermal 
fluctuations below this scale, which can be crudely taken into account using
the Gaussian distribution. We use a $\sigma_u$ corresponding to 
$\sigma_T \sim 1 500 K$. The tail of the Gaussian distribution extending
below $u_0$ can be considered as cold gas. Consequently
we can define the fraction $f_c$ of cold gas in a warm particle as

$$ f_c= 0.5(1-\hbox{erf}({u-u_0 \over \sigma_u}) ). $$

\noindent
This is the fractional mass that is used to produce cold gas particles following
the algorithm described in Sec. 3.2.

\subsubsection{Star formation}

There are two steps involved in modeling star formation. First, select 
under which conditions stars can be formed from the gas, then specify at which 
rate the
formation occurs. Katz (\cite{Katz92}) proposes a set of criteria: stars can
form only in Jeans unstable regions and within a convergent flow of the gas. 
In these regions they forms at a rate following a Schmidt law. Navarro \& White
(\cite{Navarro}) add the condition that the star forming regions are cooling
rapidly. They reduce the cooling time criterion and the Jeans instability 
criterion to a density threshold. This is coherent with results derived from
the observations (Martin \& Kennicutt \cite{Martin}). Other authors argue that
the limited resolution of the simulations, where a gas particle has usually
the mass of a whole GMC, makes the use of dynamical criteria unreliable (e.g
Mihos \& Hernquist \cite{Mihos}, Hultman \& Pharasyn \cite{Hultman}). We agree
with these considerations and use a simple Schmidt law to describe star
formation:

\begin{equation}
\qquad
{ d \rho_{\hbox{\tiny star}} \over dt} = C \rho_{\hbox{\tiny  gas}}^n\,\,.
\end{equation}

\noindent
$C$ is a constant combining a star formation efficiency, a typical formation
time and a typically density.
All cold gas particles are candidate for star formation and the
star formation rate is fixed by a Schmidt law with index $1.5$ or $1$.
Finally let us mention the work by Gerritsen \& Icke (Gerritsen \& Icke 
\cite{Gerritsen}), who solely rely on dynamical criteria such as convergent flow
and Jeans instability, and attempt to derive a Schmidt law as a result of the 
simulations.

\subsubsection{Supernovae feedback}

A lower estimate for the amount of energy released in supernovae for a Salpeter
IMF is $10^{48}$ erg per solar mass of formed stars. During the first 2 Gyr of
its life, a galaxy converts a large fraction of its gas content into stars.
The total energy released by supernovae during this process is many times larger
than the total kinetic energy of the galaxy. It appears important to take this
phenomenon into account. However since most of the feedback is in the
form of thermal energy in the dense gas environment of young stars, it will
be quickly radiated away. A small fraction is also fed back as kinetic
energy in expanding gas shells. This phenomenon is strong enough to
eject gas from dwarf galaxies. For a a detailed analysis of the different
algorithms see Thacker \& Couchman (\cite{Thacker}). In our model, we include 
both thermal and kinetic feedback. Kinetic energy is imparted to any neighboring
gas particles in the form of a velocity kick opposite to the forming star 
direction.
Thermal energy is directly deposited on a
small number of neighboring cold gas particles ($\sim 3$), which are converted
into warm gas particle and brought to a temperature of 20 000K to 100 000K 
depending on the simulation.
The thermal feedback is largely affected
by the damping of the thermal evolution of SPH particles described in 
Sec. 2.2.1.
It actually has an effect similar to imposing an adiabatic period of a few Myr
as described in Thacker \& Couchman (\cite{Thacker}). We will show in Sec. 6.1
that the thermal feedback associated to the damping has a definite effect on
the mass balance between the phases.

\subsubsection{Stellar mass-loss}

Jungwiert et al.  (\cite{Jungwiert}) show that within 10 Gyr, 
supernovae explosions and stellar winds can return over 40 \% of the stellar 
mass to the ISM for a star population following Scalo's IMF 
(Scalo \cite{Scalo}). Of these 40 \%, less than 10 \% are expelled in
supernovae explosions. Consequently, the instantaneous recycling approximation 
(Tinsley \cite{Tinsley}) used in most numerical simulations including star
formation and metallicity enrichment (e.g. Hultman \& Pharasin \cite{Hultman})
can only be a first step. Indeed, stars keep releasing gas through
stellar winds long after they have separated from the gas clouds where they
were born, and refill the gaseous disk  proportionally to the smooth star
density field, not only in locations where stars are formed. Rosen \& Bregman
(\cite{Rosen}) improved on the instantaneous recycling approximation
by using a constant mass loss rate for the stars. Jungwiert et al. 
(\cite{Jungwiert}) and Lia et al. (\cite{Lia}) propose much more detailed
models. They derive a time dependent mass loss rates for various IMF, and include
them in simulations. In our model we use a simple version of Jungwiert et al.
mass loss rate:

\begin{equation}
{d M_{\hbox{\tiny gas}} \over dt} = M_{\hbox{\tiny star}} {c \over (t-t_{\hbox{\tiny birth}}+T_0)}. 
\end{equation}

\noindent
We use the constant values $c=0.055$ and $T=5$ Myr. Our algorithm for 
mass exchange makes it unnecessary to deal the with merging of small star 
particles as in Jungwiert et al. (see Sec. 3.2).

\subsubsection{Metal enrichment}

The production of heavy elements in stars and subsequent ejection in SN 
explosions or stellar winds lead to the modification of the chemical 
composition and cooling properties of the ISM. Since our model includes both
stellar formation and stellar mass loss, it is natural to follow the 
chemical evolution of the ISM. However, as mentioned in Sec 2.2.1, we will not 
take the modifications of the cooling properties into account.

We use the yield $y$, defined as the mass of metal produced per unit mass of 
stellar hydrogen (Tinsley \cite{Tinsley}). If a star was initially formed from
gas with a metallicity $Z_s$, the gas produced by this star though the
mass loss process has a metallicity:

$$ Z_g = Z_s + y (1-Z_s)$$

During the gathering process used to produce gas particles or stars
(see Sec. 3.2), the metallicity of the different 
contributions is averaged. This introduces a numerical diffusion at the 
resolution scale. The physical processes leading to metal enrichment of the
gas are stellar winds and supernovae explosions. The diffusion of metallicity
in the actual ISM is probably realized through turbulent convective mixing.

The main difference of our model with the instantaneous
recycling approximation (Tinsley \cite{Tinsley}) is that it works within the 
continuous mass-loss scheme, so enriched gas is not only released in star 
forming regions but throughout the stellar disk.

\section{Numerical methods}

\subsection{Tree-SPH}

We use a N-body representation of the system. Gravitation and hydrodynamics are
the main factors driving the evolution. To compute the
corresponding forces acting on each particle, we use the now familiar tree-SPH
method (Hernquist \& Katz \cite{Hernquist89}). 

Gravitation forces are computed
using the so-called tree algorithm (Barnes \& Hut \cite{Barnes86}). This particular
implementation was used in previous works, for example Semelin \& Combes 
(\cite{Semelin00}). Following this algorithm, the contribution by a group of 
distant particles to the gravitational force  acting a given particle 
can be computed using the multipole moments of the group mass distribution.
It is actually computed if the angular size of the group, seen from the particle, is smaller
than an opening criterion $\theta$. Otherwise, the group is divided into
subcomponents. Throughout this work we use a value of $\theta = 0.8$. The
multipole expansions  are carried out to quadrupole moments. With this setup,
the typical error on the gravitational force is smaller than $1 \%$ 
(Hernquist \cite{Hernquist87}).

Hydrodynamic forces are computed through the SPH method (Gingold \& Monaghan 
\cite{Gingold}, Lucy \cite{Lucy}). For a complete description of the
method, see, for example, Monaghan (\cite{Monaghan92}).
However, implementations of the SPH technique come
in many different flavors (see Thacker et. al. (\cite{Thacker}) for a 
comparative study). Let us summarize our choices for this implementation.
We use the ideal gas equation of state:

$$ P=(\gamma -1) \rho u \qquad ( \gamma = {5 \over 3} ) $$

\noindent
We use a kernel based on a spherically symmetric spline function 
(Monaghan \& Latanzio \cite{Monaghan85}). The individual smoothing length
of each particle is adaptative and we use the arithmetic average 
$h_{ij}={h_i+h_j \over 2}$ to compute the interaction between two particles.
We have taken care to insure that whenever particle B acts as a 
SPH-neighbor of particle A, then the reverse is true, particle A acts as a 
SPH-neighbor of particle B. This is important for a good energy conservation.
The time update of the individual smoothing
length follows the scheme proposed by Hernquist \& Katz (\cite{Hernquist89}).
Finally, we use the viscosity described in Monaghan (\cite{Monaghan92}). 

To measure the quality of our implementation, we run a simulation that is 
becoming a standard validation test for this type of codes: the collapse
of an initially static, isothermal sphere of self-gravitating
gas. This was first studied by Evrard (\cite{Evrard}) and has been
reproduced by many authors since then 
( e. g., Hernquist \& Katz \cite{Hernquist89}, 
Thacker et al. \cite{Thacker}, Springel et al.
\cite{Springel}...). Let us consider a sphere of radius $R$, mass $M$,
density profile

$$ \rho(r)= {M \over 2 \pi R^2} {1 \over r} \,\, , $$

\noindent
and uniform internal energy

$$ u= 0.05 {G M \over R}\, . $$

The sphere is composed of $N$ particles initially at rest, distributed
within $R$ with the proper probability to realize the profile $\rho(r)$.
We use $R=1$, $M=1$ and $G=1$. Since our implementation
uses a single time step, we have to choose a small value. We set 
$\delta t=0.001$, and integrate from $t=0$ to $t=3$. We use $N=10000$ particles,
a softening $\epsilon=0.02$ and an opening criterion $\theta=0.8$.
The evolutions of total, kinetic, potential and internal energies are given
on Fig 2. The maximum relative variation of the total energy is 0.3 \%. The
evolutions on Fig 2. match closely those given by Thacker et al. (\cite{Thacker})
or Springel et al. (\cite{Springel}). This result was obtained using 40 
neighbors for each SPH particle. We have observed that using a smaller number
of neighbors, 15, for a constant $N$, produces a similar evolution with a
thermal peak 35 \% higher, just as larger $N$ produce higher peaks. This 
boils down to the degree of smoothing of the central collapse by the SPH 
algorithm, the smaller the smoothing length, the larger the thermal peak.

\begin{figure}[t]
 \resizebox{\hsize}{!}{\includegraphics{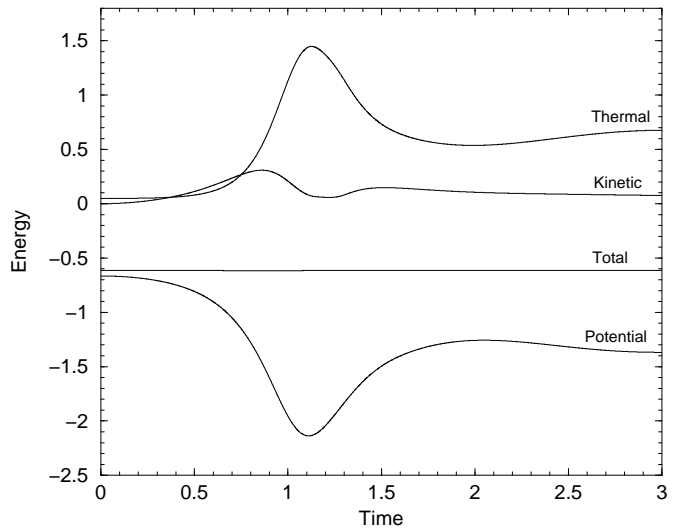}}
\caption{Evolution of total, thermal, potential and kinetic energies of an
initially isothermal sphere of gas with density profile $\rho(r) \sim {1 \over r}$. We use dimensionless unit, and $G =1 $.
The thermal and potential extrema correspond to the instant of maximal collapse.}
\end{figure}

\subsection{ Implementation of mass exchanges}

As presented in sec. 2, our model takes into account four different types of
particles: dark matter, stars, warm gas and cold gas. As a result from cooling,
warm gas can be converted to cold gas. In dense regions, cold gas is then
converted to star following the star formation rate fixed by a Schmidt law. In
turn, stars expel warm gas following the mass-loss prescription (eq. 4). 
And finally,
cold gas can be transformed into warm gas as a result of supernovae feedback.
Except for the vaporization of cold gas to warm gas, which receive a simple treatment, we use a common numerical procedure to realize these exchanges.

A number of procedures have been used in the literature, applied mostly to
star formation. In one way or another, all procedures use a variable to
keep track of the fractional mass of stars already produced in a given gas 
particle (this can be generalized to any two phases). 
We can then separate procedures in two types. In the first type, pioneered
by Katz (\cite{Katz92}), the gas particle spawns small star particles each
time its fractional star-mass reaches a chosen low threshold value. In this procedure
the dynamics of collisionless star particles is differentiated early-on
from the dynamics of the gas. The drawback is that it leads to the
creation of many star particles with small masses. This increases the CPU cost
and leads to two-body relaxation problems: the young stellar disk, composed
of low-mass particles will be sensitive to the dynamical heating by heavier
particles. We will show in sec. 5 that this is a sensitive issue, especially 
with heavy dark-matter particles. Jungwiert et al. (\cite{Jungwiert}) have
applied this procedure to the mass-loss of the stars. They have taken
care of some of the drawbacks by including schemes to merge small-mass
particles of the same type. 

In the second type of procedure, the entire gas
particle is converted to star when the fractional star-mass reaches a high
threshold, 80\% for example. This was used by Mihos \& Hernquist (\cite{Mihos}).
They also choose to subtract the fractional star-mass from SPH density
evaluations to avoid having a 70 \% star particle behave as a full gas 
particle. Other authors choose not to, e.g. Thacker \& Couchman 
(\cite{Thacker00}). An extension of this procedure is to use a probabilistic
approach to convert the gas particle into a star particle, with a probability
function depending on the fractional star-mass 
(see, e.g., Lia et al. \cite{Lia}). The main drawback
is that large amounts of star-mass may have to
follow gas-particle trajectories for a long time.

Our scheme, similar to Steinmetz \& Muller (\cite{Steinmetz}), combines the advantages of keeping the number of particle constant
(no small mass particles are produced) and keeping the fractional mass of
stars trapped in gas particle very small. We simply use an SPH-like gathering
procedure. When the fraction of star-mass in a gas particle is higher than
5 \%, we examine the star-mass content of its SPH neighbors. If there
is enough star-mass in the neighboring particles  to make one full star 
particle, the 
mass is gathered in the initial particle, which is then converted into a star.
Using 15 SPH neighbors, the average star content of a gas particle is
smaller than 10\%. This procedure is also applied to convert warm gas into
cold gas, and stars into warm gas. Consequently we have to keep track of
a ``smoothing length'' for all three types of particles.

\section{Formation of a stellar disk: reference case}

\subsection{Initial conditions}

The dark matter is distributed in a halo with a Plummer density profile:

$$ \rho_{dm}(r)={ 3 M_{dm} \over 4 \pi r_{dm}^3 } ( 1 + {r^2 \over r_{dm}^2})^{-5/2}. $$

We set $r_{dm} = 5 $kpc, and $M_{dm}=1.7 10^{11} M_{\odot}$.
We introduce a cut-off at $5 r_{dm}$.  

The baryonic matter (warm gas at t=0) is distributed in a rotationally supported
thin disk defined by a Miyamoto-Nagai density profile 
(Miyamoto \& Nagai \cite{Miyamoto}):

$$ \rho_{bm}(r,z)= ( {z_{bm}^2 M_{bm} \over 4 \pi }) { r_{bm} r^2 +( r_{bm}
+ 3 \bar{z})(r_{bm}+\bar{z})^2 \over
[ r^2 + (r_{bm}+\bar{z})^2]^{5/2}\bar{z}^3 }
$$

\noindent
where,

$$ \bar{z}=\sqrt{z^2+z_{bm}^2}. $$

\noindent
$r_{bm}$ is the typical radius of the disk and $z_{dm}$ is the typical vertical
thickness. We use $r_{bm}= 5$ kpc and $z_{bm}=0.3$ kpc, with cutoffs at 
$3 r_{bm}$ and $3 z_{bm}$. The total baryonic mass is $M_{bm}= 5.7 \,\,10^{10}
M_{\odot}$ ($1/3$ of the dark matter halo mass). 
The angular speed $\Omega(r)$ for the baryonic particle is derived from the
density profiles of dark matter and baryonic matter for global rotational 
support. Velocity dispersions are derived from disk stability criterion:

$$ \sigma_r \sim {3.36 G \Sigma \over \kappa } \qquad \sigma_{\theta} \sim
\sigma_r {\kappa \over 2 \Omega} \qquad \sigma_z \sim \sqrt{G z_{bm} \Sigma}, $$

\noindent
where $\Sigma$ is the surface density of the disk, and $\kappa$ the epicyclic
frequency. Exact values are chosen to produce a Toomre stability criterion 
$Q \sim 5 $ in the center, falling off to $ Q \sim 2 $ at the edge of the disk.

Initially, all baryonic particles are SPH gas particle with a temperature of 
$ 15 000 $K, and metallicity $Z=0$.

For the reference simulation we use the following parameters.
A Schmidt law with exponent 1.5 is used for star formation. The number of
baryonic particle is $50 000$, the number of dark matter particle is $10 000$.
This makes dark matter particles 15 times heavier than baryonic particles. The
gravitational smoothing is $ 30$  pc for baryonic matter and $300$ pc for
dark matter. The minimal SPH smoothing is accordingly set to $30$ pc. The time
step is 1 Myr and the integration is carried out over 2 Gyr.

\subsection{results}

\begin{figure*}[t]
\centering
\includegraphics[width=17cm]{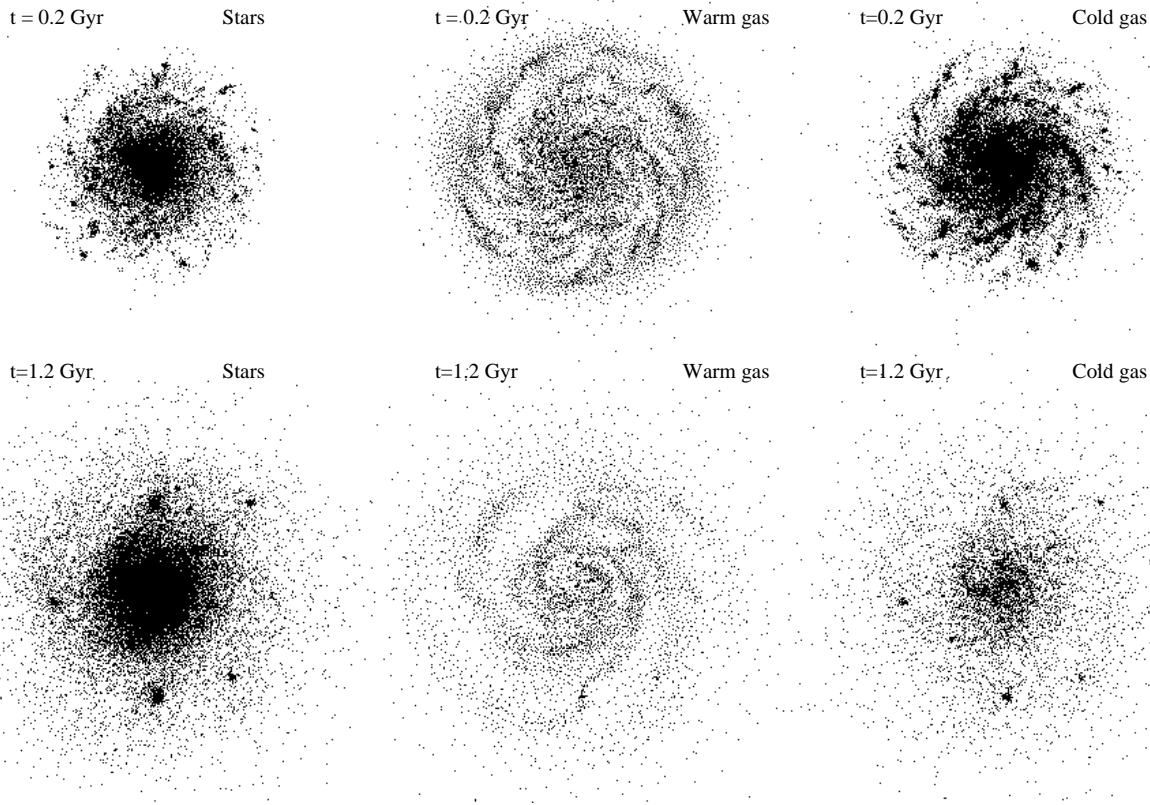}
\caption{Reference simulation. Configuration of the three baryonic phases after 200 Myr (top) and
after 1.2 Gyr (bottom).  }

\end{figure*}

Face-on views of the various baryonic phases are presented in Fig. 3. The
snapshots are taken at 0.2 Gyr and 1.2 Gyr. They show the formation of the stellar
disk and the depletion of the gas phases.
One interesting point in the stellar disk formation is that stellar clumps form
in the outer, most unstable regions of the disk during the first few
hundred Myr. They subsequently merge in the central part of the disk. 
We have encountered this phenomenon for a wide choice of parameters. Only with
very high initial values of Toomre criterion ($Q \sim 8$) throughout the disk
does it disappear. It is physically unlikely that such a $Q=8$ disk
is formed before any star formation takes place, since dissipation tends to
decrease Toomre criterion in the gaseous disk.
Noguchi (\cite{Noguchi}) also forms clumps for a cold collisional
disk of gas and suggests that it can be a mechanism for the formation of a bulge.

Another morphological aspect worth mentioning is the behavior of the spiral
structure. It is strong in the three phases during the first few 100 Myr. It then
fades, especially in the stellar disk. Possible causes for this
fading are the absence of external perturbations to drive the spiral
structure, and the lack of resolution leading to unphysical heating of various
baryonic components (see also Sec. 5)

\begin{figure}
 \resizebox{\hsize}{!}{\includegraphics{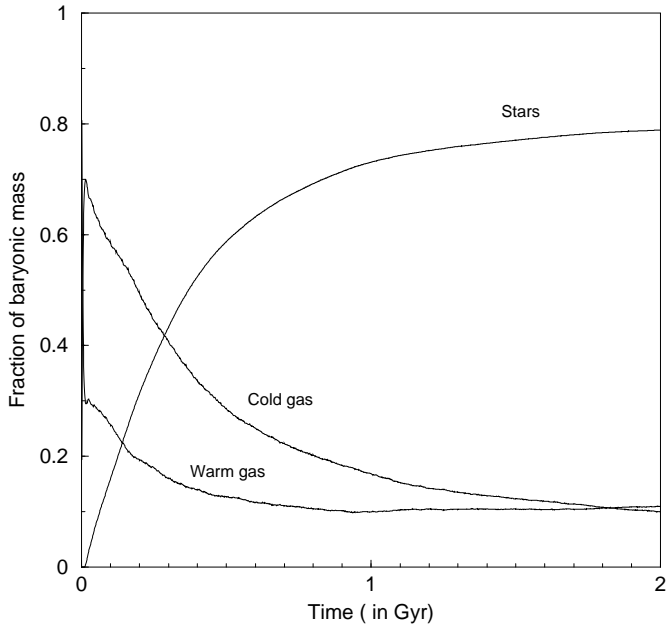}}
\caption{Evolution of the composition of the baryonic matter. The mass fraction
of each of the three components is plotted against time.}
\end{figure}

\begin{figure}
 \resizebox{\hsize}{!}{\includegraphics{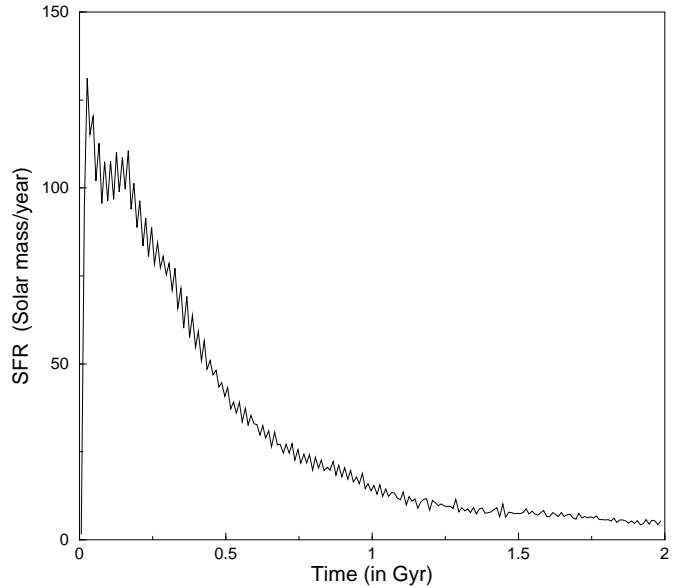}}
 \caption{ Star formation rate in the reference simulation, as a function of 
 time. The star formation law is a Schmidt law with exponent 1.5.}
 \end{figure}

Fig. 4 shows the time-evolution of the mass of each baryonic phase.
A large part of the warm gas is almost instantaneously turned into cold gas,
as thermal equilibrium is reached within a few time steps. Then mass exchanges 
proceed smoothly.
Within 500 Myr, the warm gas mass reaches a quasi-constant value, as the balance
between the gain from stellar mass loss and the loss from cooling into cold gas
is reached. This is a self-regulated equilibrium: if the warm gas mass is
depleted through cooling, its density falls and the cooling becomes less 
efficient. If it increases through stellar mass loss, the density rises, the
cooling becomes more efficient, increasing the rate of transfer to the cold gas
phase. On long time scales (several Gyr), the warm gas mass would eventually
decrease as stars grow older and release less warm gas. The cold gas phase is
continuously depleted although the process is slow after 2 Gyr. We will examine
in Sec 6 the influence of various parameters on the mass evolution of 
the phases. Fig. 5 shows the evolution of the star formation rate. 
The rate is high during the first Gyr, during which 75\% of the gas is turned 
into stars. It decreases steadily with time. This is not fully into accord with
the observations, which suggest more constant star formation rate. The missing
factor here is probably the accretion of gas, which would replenish the gas
disk and sustain star formation rate.

\begin{figure}
\resizebox{\hsize}{!}{\includegraphics{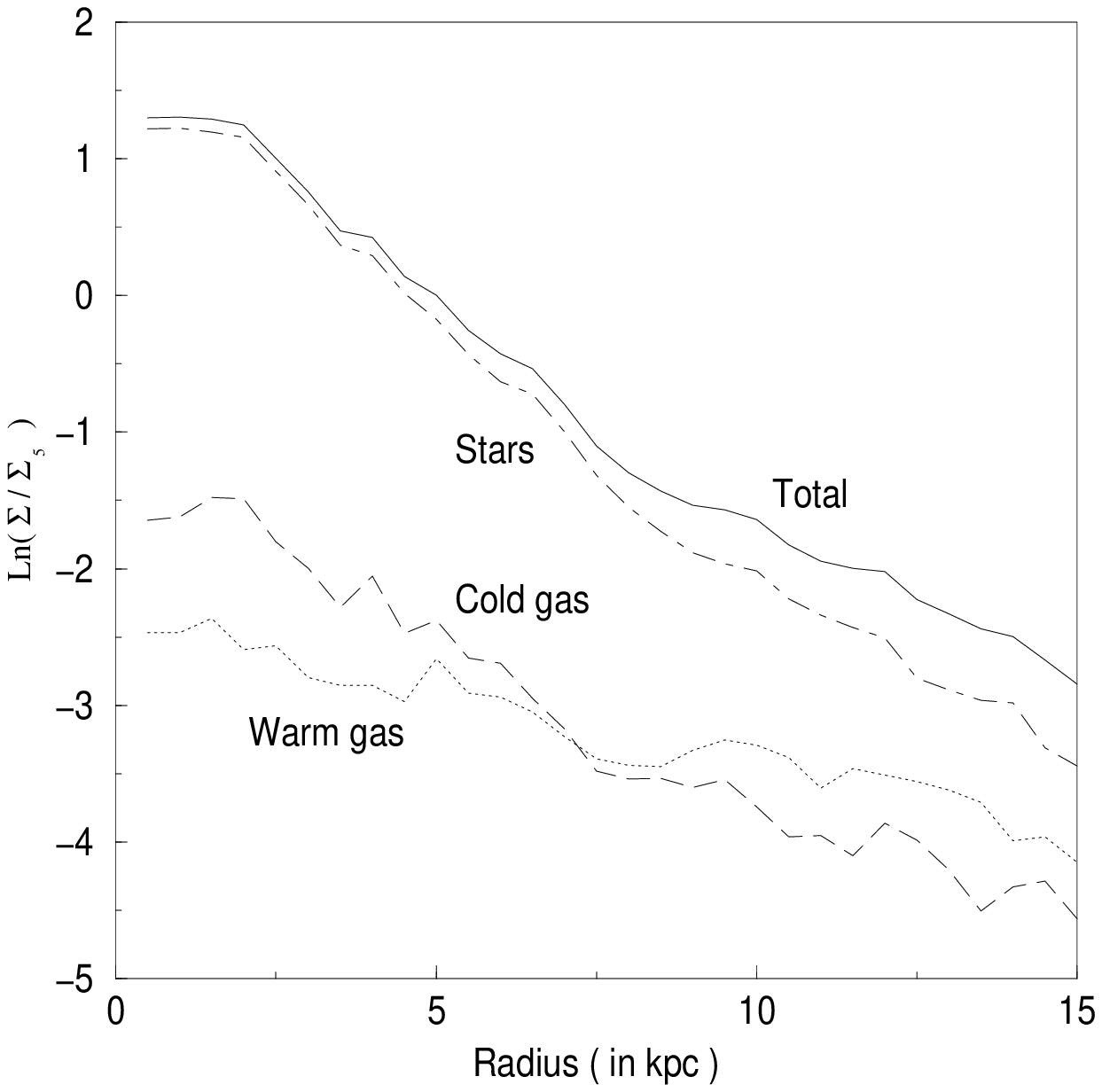}}
\caption{ Surface density profiles of the various baryonic components of the
disc as functions of the radius. The density scale is logarithmic. The profile
are roughly exponential with 10 kpc typical radius for the warm gas, and 2 kpc
for the stars.}
\end{figure}

\begin{figure}
\resizebox{\hsize}{!}{\includegraphics{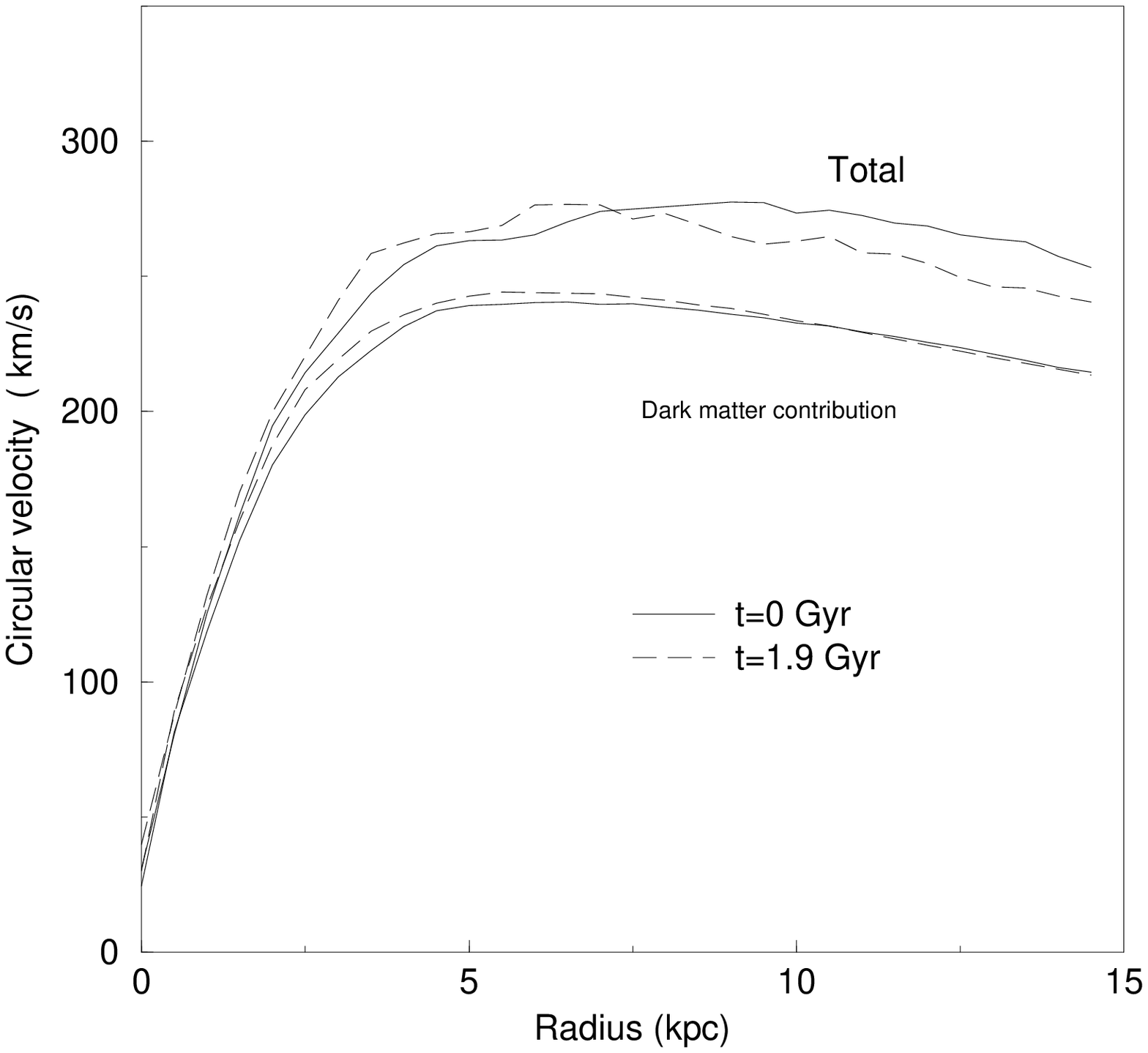}}
\caption{ Rotation curve of the disc (circular velocity as a function of radius)
at the beginning and the end of the reference simulation.
The curve shows little evolution and is rather flat in the
outer part of the disk. The dark matter contribution is large, producing a stable disk.}
\end{figure}

The surface density profiles of each baryonic phase at 2 Gyr are plotted  on 
Fig. 6. The stellar disk exhibits an exponential profile with a 2 kpc typical
radius. The warm gas disk profile is also reasonably fitted by an exponential,
with a rather larger typical radius of $\sim 10$ kpc. The cold gas disk falls
in between. Finally, Fig. 7 shows that the circular velocity profile exhibit
little evolution during the first 2 Gyr of the simulation. The profile is rather
flat at $r > 5$ kpc, with a typical velocity of $\sim 250$ km s$^{-1}$.

\subsection{ Modification of the parameters}

\begin{table*}
\centering
\begin{tabular}{|l||l|}
\hline
& \hskip 1cm Differences with model 1 \\
\hline
\hline
\hskip 1cm Model 1 \hskip 1cm \hbox{}& \hskip 1cm Reference model (see Sec. 4)\\
\hline
\hskip 1cm  Model 2 \hskip 1cm \hbox{}& \hskip 1cm Unique gravitational softening (30 pc) \\
\hline
\hskip 1cm Model 3 \hskip 1cm \hbox{}& \hskip 1cm Analytical dark matter halo\\
\hline
\hskip 1cm Model 4 \hskip 1cm \hbox{}& \hskip 1cm High resolution simulation: 500 000 particles \hskip 0.5 cm\\
\hline
\hskip 1cm Model 5 \hskip 1cm \hbox{}& \hskip 1cm Schmidt law with exponent 1.\\
\hline
\hskip 1cm Model 6 \hskip 1cm \hbox{}& \hskip 1cm No stellar massloss\\
\hline
\hskip 1cm Model 7 \hskip 1cm \hbox{}& \hskip 1cm No supernovae feedback\\
\hline
\hskip 1cm Model 8 \hskip 1cm \hbox{}& \hskip 1cm Purely thermal supernovae feedback\\
\hline
\hskip 1cm Model 9 \hskip 1cm \hbox{}& \hskip 1cm High thermal supernovae feedback\\
\hline
\hskip 1cm Model 10 \hskip 1cm \hbox{}& \hskip 1cm Lighter dark matter halo\\
\hline

\end{tabular}
\caption{ Description of the various models used in the simulations.}
\end{table*}

The relative complexity of the model that we propose makes it impossible to
study the action of all the parameters independently. Some parameters have been
tuned to produce physically acceptable results, and remain unchanged in our
study. This is the case, for example, of the rate of dissipation through
inelastic collisions in the cold gas: it is tuned so that cold gas looses a
few percents of its velocity dispersion every 100 Myr. Another example is the
temperature of the phase transition between warm and cold gas, set at 11000 K.
Changing the value would tilt the mass-equilibrium between the phases. We have
tuned it in connection with other parameters such as the heating rate $\Gamma$
to produce similar masses for the two components after 2 Gyr in Model 1. 

The modifications we have chosen to study define 10 models described in Fig. 8.
Model 1 is the reference model described in Sec. 4. In model 2, heavy
dark matter particles have the same gravitational softening as light baryonic
particles. In model 3, the dark matter halo is not realized with particles,
but using the static analytic profile given in Sec. 4.2. Model 4 includes
400 000 baryonic particles and 100 000 dark matter particles with a unique
$15$ pc gravitational softening, the time step is reduced to 0.5 Myr. In model
5, a Schmidt law with exponent 1 instead of 1.5 is used to compute the star formation rate.
In model 6, the stellar mass loss process is simply shut off, and an 
instantaneous recycling approximation is used for metal enrichment.
In model 7, supernovae feedback is switched off. In model 8,
the kinetic part of the supernovae feedback is switched off. In model 9, the
supernovae feedback is purely thermal again, but gas particle are heated up
to 100 000 K instead of 20 000 K. Last, in model 10, the mass of the
dark matter halo is three times smaller than in the reference model.

\section{ Disk thickness: the numerical heating}

Numerical simulations of galactic disks usually focus on the baryonic matter
behavior which can be easily compared to observations. It is then tempting
for a given CPU cost to increase the scale resolution of baryonic matter
at the expenses of the dark matter, using many light baryonic matter
particles and few heavy dark matter particles. This is the case in most of
the simulations in this work, where the mass ratio between dark matter and
baryonic matter particles is between 5 and 15. This raises the question of how to
handle two-body relaxation within each component, and more importantly
between the 2 components. The two-body relaxation time-scale is usually
controlled by the value of the gravitational smoothing length. There is 
actually an optimal value which limits both two-body relaxation and the bias
in the force evaluation (see, e.g. Romeo \cite{Romeo}, Athanassoula 
\cite{Athanassoula}). How to deal with several particles species with different
masses is not as well established. The main lead is to use variable softening
parameters. Dehnen (\cite{Dehnen}) presents an analytic study of the effect
of using variable smoothing length
on the bias in the force computation. Binney \& Knebe (\cite{Binney})
present a numerical study including two species of particles with different
masses in a cosmological context, and find that using different softening
values can reduce mass segregation, an effect arising from two-body relaxation
between the different species. 

Our early simulations showed a thickening of the stellar disk.
It was quickly realized that the heavy dark matter particles were 
responsible for the heating of the baryon disk. Being organized in a spherical halo, they have a
large velocity dispersion perpendicular to the disk which is transmitted to
the baryonic particles. We have studied the importance of this effect under
various simulation settings.

\begin{figure}[t]
\resizebox{\hsize}{!}{\includegraphics{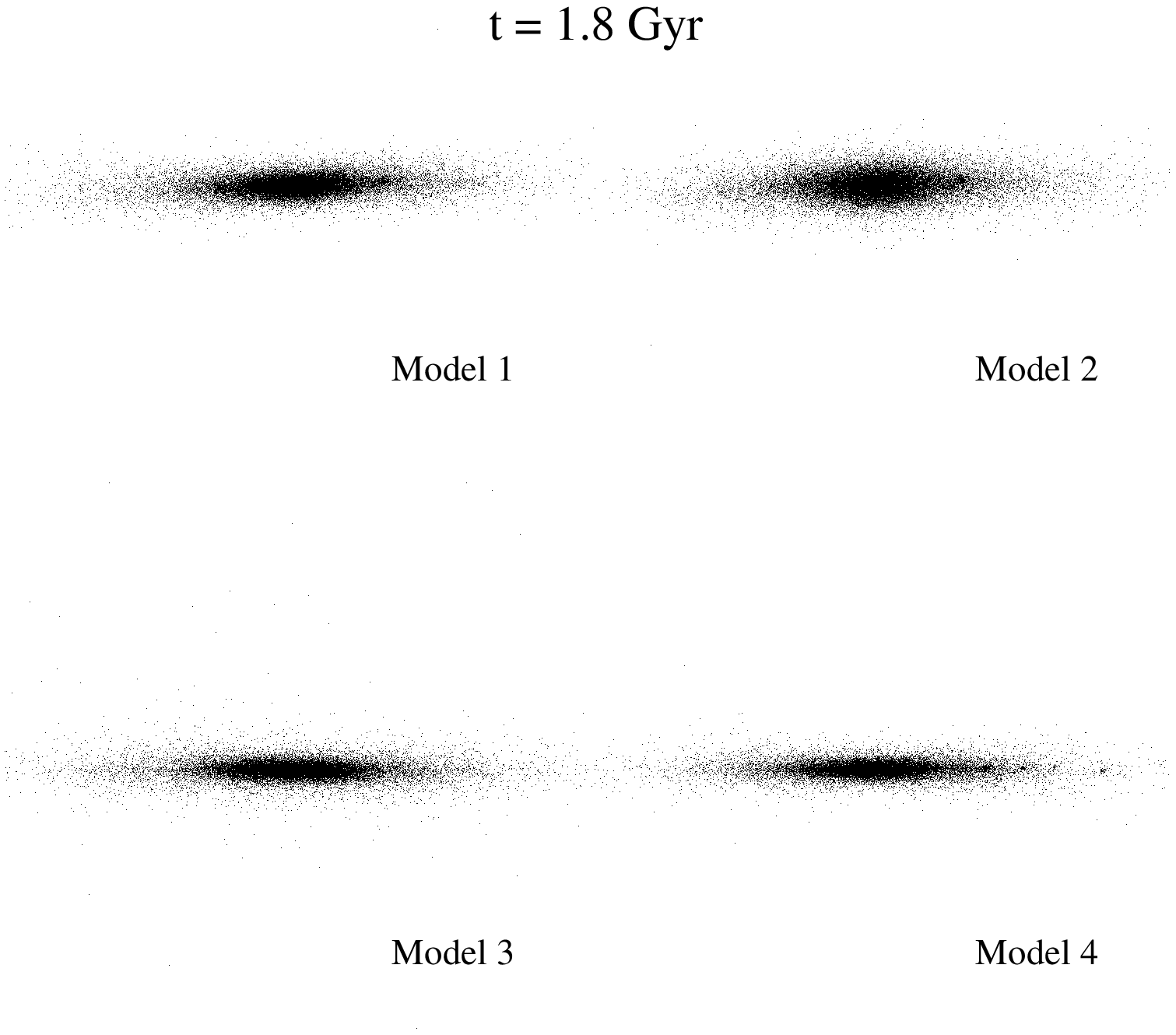}}
\caption{Edge-on views of the stellar component of the disk of different models,1.8 Gyr after identical initial configurations. Models exhibit different degree
of thickening.
}
\end{figure}

\begin{figure}[t]
\resizebox{\hsize}{!}{\includegraphics{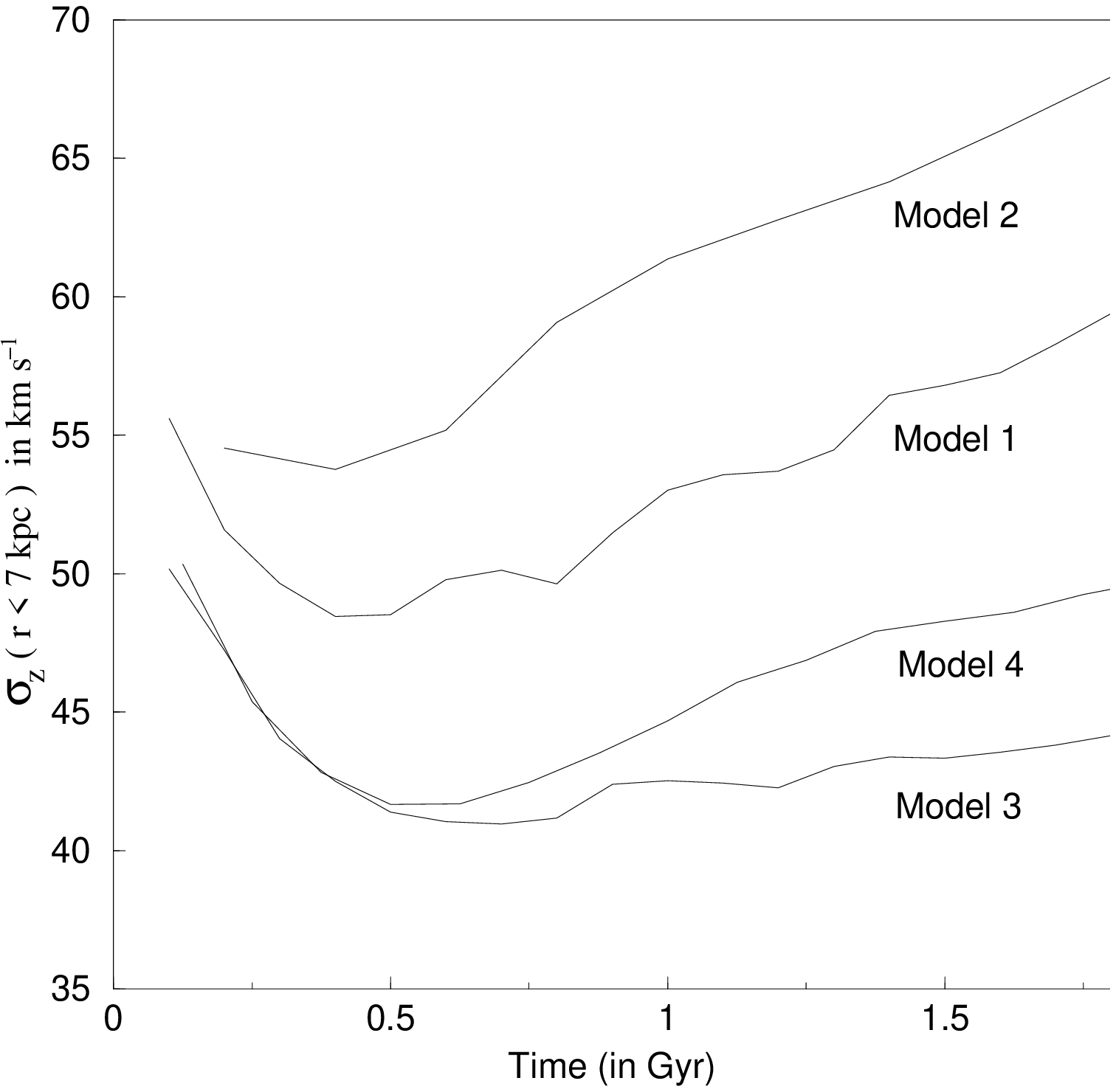}}
\caption{The vertical velocity dispersion of the stellar disc is averaged in
the inner part of the disc ( $r < 7$ kpc). Its evolution is plotted for different models, tracking the thickening of the disk.}
\end{figure}

The thickening of the stellar disk was compared for models 1, 2, 3 and 4.
Fig. 9 shows the aspect of the disk seen edge on after 1.8 Gyr. Model 2, with
a constant softening length of 30 pc produces the thickest disk. Model 1, with
softening 30 pc for the baryonic matter and 300 pc for the dark matter, 
produces a  thiner disk, but still quite thicker than models 3 and 4. Model 3,
with an analytic halo component removes all heating of the disk due to the dark
matter particles, thus providing a reference point. Interestingly model 4, including
500 000 particles also produces a very thin disk. The relaxation time has increased enough
with the number of particles to produce little thickening within 2 Gyr.

Fig. 10 gives a quantitative evaluation of the phenomenon. We plot the 
average vertical velocity dispersion of the baryonic matter within a
radius of 7 kpc as a function of time for the four models described above.
Initially the matter is mainly in the form of dissipative gas, which explains
the initial decrease of the velocity dispersion. Conclusions pertaining to
the heating by dark matter can be drawn from the latter times ( $>$ 0.5 Gyr), when
the disk has settled into a quiet evolution. The conclusions are the same as
drawn from Fig. 9. The only new element is that, in model 4, the heating
is not completely suppressed, which was not obvious in Fig. 9.

Our main conclusion is that, while different smoothing length helps reduce
the transfer of vertical momentum between dark matter and baryonic matter,
a greater number of particle is the best solution.

\section{ Sensitivity of physical quantities to parameter variations}

\subsection{ Mass equilibrium between the phases}

As explained in Sec. 4.4 and shown on Fig. 4, the parameters of the various 
physical processes have been chosen in the reference simulation to
produce a mass equilibrium between the phases that is only slowly evolving
after 2 Gyr. The gas fraction of the baryonic mass is then about 21 \% and 
slowly decreasing, which is compatible with the value of $\sim 10 \%$ 
observed in present days, $\sim 10$ Gyr old galaxies. 

Most of the physical processes included in the model can affect the mass 
balance between the phases. We have studied the effect of varying some of the
parameters involved in the processes modeling. The result are summarized in
Fig. 11. The mass fraction of each baryonic component is given at 0.5 Gyr
, which marks the end of the initial most unstable period., and at 2. Gyr, the
end of the simulations. 

Model 5 shows the effect of changing the exponent from
1.5 to 1. in the Schmidt law. The constant $C$ in eq. 3 is adjusted to yield
similar overall star production; Model 5 produces less stars in the initial
period of high gas density, and more in the later period of low gas density. The
final fraction of cold gas is smaller in Model 5 than in the reference model,
more of it having been turned into stars. The warm gas fraction is barely
affected. 

Model 6 is a run where stellar mass loss has been switched off.
It is interesting to notice that, although stars eject warm gas particle, it
is the cold phase that is mostly depleted when mass loss is switched off. We
interpret this as follows. The warm gas equilibrium mass which appears on 
Fig. 4, is associated with a critical density fixed by the cooling process.
If the gas falls below this density, transfer to the cold gas phase almost 
stops. The injection of mass from stellar mass loss does barely increases the
density above the
critical density since the cooling increases in proportion to the density. 
Its main effect is actually to maintain a mass-flow from stars to warm 
gas to cold gas. If this flow  dies up at the source, by switching
off stellar massloss, the warm
gas settles at its critical density and stops producing cold gas. Cold gas is
then steadily depleted through star formation.

Models 7, 8 and 9 study the effect of supernovae feedback. From the result of
model 7, where SN feedback is absent, we assert that SN feedback does not
affect the average star production, but that it has a strong influence on
the balance between cold and warm gas. We can guess that the evaporation
of cold gas by thermal input from SN, associated with the damping of thermal
evolution is a more important factor in this regard than the kinetic feedback.
This is checked in model 8, where SN feedback is purely thermal. The fraction
of each baryonic component is very similar to the reference model. Model 9
shows that using a high thermal feedback does not change the balance much more.

The largest variation in this study is from 5.1 \% to 17.6 \% of warm gas 
at 2 Gyr between models 6 and 7. The largest variation with respect to the 
reference model is from 17.6 \% to 10.2 \% of warm gas at 2 Gyr. These are
limited variations which point to the stability of the model as far as the
composition of the baryonic matter is concerned.

\begin{table*}
\centering
{\large
\begin{tabular}{|c||c|c|c|c|c|c|}
\hline 
Model & \multicolumn{3}{|c|}{ At 0.5 Gyr} & \multicolumn{3}{|c|}{ At 2. Gyr} \\
\hline
 & Stars & Warm gas & Cold Gas & Stars & Warm gas & Cold gas \\
\hline
1 & 58.7 \% & 12.5 \% & 28.8 \% & 78.9 \% & 10.9 \% & 10.2 \% \\
\hline
5 & 53.7 \% & 13.3 \% & 33 \% & 84.8 \% & 10.3 \% & 4.9 \% \\
\hline
6 & 65.4 \% & 11.8 \% & 22.8 \% & 86.2 \% & 8.7 \% & 5.1 \% \\
\hline
7 & 58 \% & 7.6 \% & 34.4 \% & 77.5 \% & 4.9 \% & 17.6 \% \\
\hline
8 & 61.2 \% & 11.3 \% & 27.5 \% & 79.5 \% & 10.4 \% & 10.1 \% \\
\hline
9 & 61.4 \% & 13.2 \% & 25.4 \% & 78.5 \% & 11.2 \% & 10.3 \% \\
\hline
\end{tabular}
}
\caption{ Composition of the baryonic matter at 500 Myr and 2 Gyr in different
models. See Fig. 8 for properties of the models.}
\end{table*}

\subsection{Metallicity profiles}

\begin{figure}[t]
\resizebox{\hsize}{!}{\includegraphics{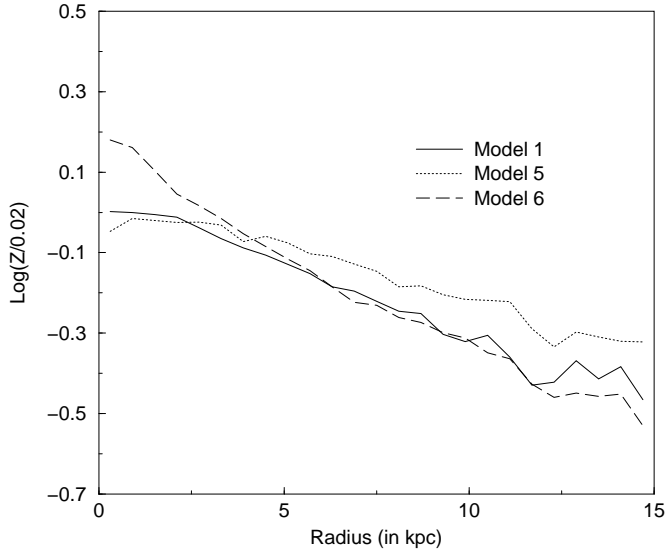}}
\caption{ Metallicity (log scale) as a function of radius for different models
(see Fig. 8). All profiles are roughly exponential.}
\end{figure}

We establish metallicity profiles at 2 Gyr to test the reliability
of the overall star formation process. Let us recall that metal enrichment
is included in the simulation only as a diagnosis, without any influence
on the cooling of the gas (see Sec. 2.2.1 for the justification). The 
metallicity
profiles of the gas component for models 1, 5 and 6 are presented in Fig. 12.
All three roughly fit an exponential profile. As can be expected the metallicity
gradient is steeper in models 1 and 6, with a Schmidt law of exponent 1.5,
than in model 5, with a Schmidt law of exponent 1. In Model 6,  stellar
mass loss is switched off, and metals are returned to the gas using the
instantaneous recycling approximation. For every unit mass of star formed,
0.4 unit mass of gas is instantaneously enriched using a yield $y=0.02$. The
difference with the reference simulation shows at small radii, where
instantaneous recycling produces a steeper metallicity gradient. Indeed this
is the region where star formation is the most active, and the instantaneous
recycling, which does not allow enriched gas to be carried  out of star 
forming regions, may overestimate the process of enrichment.

\subsection{Bar formation in a light dark matter halo}

In the reference simulation, we use a dark matter halo three times heavier than
the baryonic disk within a 15 kpc radius. This kind of 
configuration produces a stable disk and a rather quiet evolution which are 
convenient to test the model. However, it prevents the formation in the disk
of bars and grand design spiral structures, in particular $m=2$ modes. Such features appear
when the disk is less stable, that is for example if the dark matter halo is
lighter or less concentrated. Moreover, disk rotation curves derived from
observations tend to show that the contribution of the dark matter to the
circular velocity in the inner part of the optical disk is small for typical spiral galaxies (e.g. Sofue \& Rubin \cite{Sofue01}).

\begin{figure}[t]
\resizebox{\hsize}{!}{\includegraphics{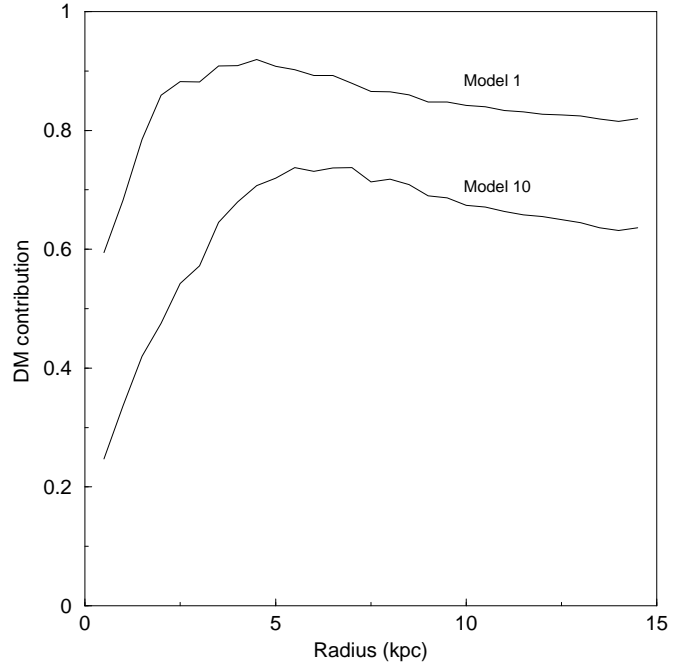}}
\caption{ Relative contribution of the dark matter halo to the disc circular 
velocity as a function of radius (see main body for exact definition).
In model 10 (light dark matter halo), the 
inner part of the disc is dominated by baryonic matter.} 
\end{figure}

In model 10, we set the mass of the dark matter halo equal to the mass of
the baryonic disk, that is three times lighter than in the reference simulation.
All other parameters are unchanged. The circular velocity of the disk in the
outer flat part of the rotation curve is $\sim 200$ km s$^{-1}$ in model 10,
and $\sim 250$ km s$^{-1}$ in the reference simulation.
Fig. 13 shows the relative contribution of
the dark matter halo to the circular velocity in the disk at $t=0 $, for
models 1 and 10. The quantity plotted is:

$$ {\hbox{\bf v}(\rho_{\hbox{\tiny DM}}) \over \hbox{\bf v}(\rho_{\hbox{\tiny tot}})}\,\, ,$$

\noindent
the ratio between the circular velocity which would result from dark matter
alone and the actual circular velocity.
We can check that, in the inner part of the disk, model 10 is
closer to reality, with a small relative contribution from the dark matter.

\begin{figure*}[t]
\resizebox{\hsize}{!}{\includegraphics{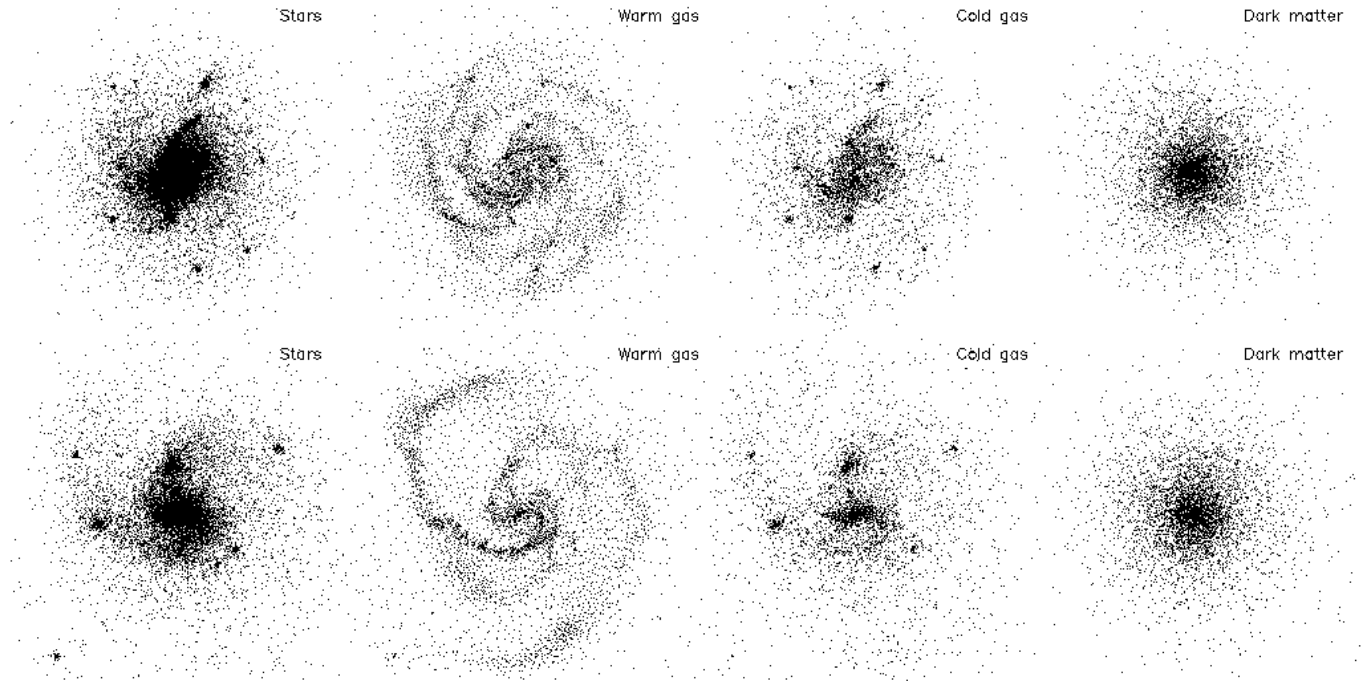}}
\caption{ Face-on plots of the four type of matter at 900 Myr for the reference
model (top) and for model 10 (bottom).}
\end{figure*}

The analysis of the evolution of the baryonic matter composition doesn't show 
any qualitative difference between model 1 and 10.
Fig 14 shows the configuration of the various components at $t=900$ Myr, for 
the reference model and for model 10. The existence of a bar and a $m=2$ spiral
mode is obvious in model 10, especially for the warm gas component. Simulation
plots show that the bar is actually present during several 100 Myr.

\section{Conclusions}

The first motivation for this work is the recognition that the cold gas
present in the ISM has morphological and dynamical properties very different
from those of the warm diffuse gas. Semelin and Combes (2000) and Semelin and
Combes (2002) explore some of the properties of the cold gas and its effect
on the galaxy formation process. Those were local studies, where the small
scale fractal structure of the gas is fully taken into account. In the
present work we try, using a simple model, to include cold gas physics in 
a global model of galaxy formation and evolution. Ideally, what we learn from
local studies of cold gas physics could be included as "sub-resolution" 
processes in global simulations. This interplay will hopefully develop in the
future.

This work presents a multiphase numerical model for galaxy formation and 
evolution. The model takes into account four different types of matter
obeying different dynamics: dark matter and stars, warm gas and cold gas.
Dark matter and star particles obey gravity only, computed in a tree algorithm.
Warm gas particles follow self-gravitating hydrodynamics implemented by
a tree-SPH algorithm, and cold gas obeys gravity and undergo inelastic collisions
(sticky particle scheme). Warm and cold gas are treated as two different 
fluids, and this is one of the innovations in the present work. In galaxies, gas and stars  exchange matter and energy through a
number of processes. The model takes into account the following: heating
and cooling of the gas, star formation, stellar mass loss, kinetic and thermal
feedback from SN explosions, metallic enrichment. We use a continuous massloss
model which goes beyond the usual instantaneous recycling approximation and
has not been integrated in such a multiphase model before.

Sec. 4 follows the evolution over 2 Gyr of an initially pure gas disk 
within a dark matter halo. The resulting galaxy exhibits properties similar
to those of observed Milky Way like galaxies. We present in particular the
evolution of the composition of baryonic matter, the star formation history,
the surface density profile of the various components and the initial and
final rotation curves. These quantities (except for star formation history) 
is reasonably similar to observed values.

The formation of a stellar disk from a pure gas component gives rise to
a large number of self-gravitating clumps; those heat the disk, merge,
and are driven towards the center through dynamical friction, where they 
contribute to form a bulge.  The cold gas phase is always more concentrated
than the warm gas phase. Although the continuous stellar mass loss replenishes
the disk in gas along its evolution, the star formation rate decreases 
exponentially with a short characteristic time-scale of the order of
1 Gyr, and therefore gas infall would be required to reproduce the observed 
star formation history in giant galaxies like the Milky Way.

Sec. 5 details the problem of the thickening of the stellar disk during
the 2 Gyr of the simulation. This thickening is due to the
two-body relaxation between star and dark matter particles. It is 
shown that using
different, larger, smoothing length  for the heavier dark matter particles
reduces the transfer of velocity dispersion, but  a better result is
obtained by using an analytic halo or by increasing the number of particles
to 500 000. We emphasize this point since many simulations of galaxy
evolution implementing complex models (including ours)
are still limited to a few $10^4$ particles for systematic studies.

Sec. 6 explores the effect of several modifications to the model. We
first study how the evolution of the baryonic matter composition is modified
under several type of circumstances, such as a different star formation law,
different stellar massloss, or different SN feedback. The model is 
rather stable in this respect. The mass composition responds to these
changes, but within reasonable limits. Constraining the final baryonic matter
composition from observations allows to discriminate between the various
possible processes. The metallicity  profiles are also computed for several
models and are found to be exponential, which validates the star formation process.

 If we try to estimate which of the processes and parameters studied
are the most significant for the overall evolution of the galaxy, it appears 
that taking into consideration a proper massloss scheme for the stars comes 
first. Indeed
it affects noticeably the composition of baryonic matter, it affects the
metallicity profiles, and, although it was not studied in detail in this work,
it may affect the morphology of the galaxy by changing the sites of star formation. Next, comes the specific star formation law, which is of course
central to reproduce the star formation history. And third, the existence
of SN feedback which allow to sustain a warm gas phase. It is an interesting
point that the specific strength of the feedback is not a sensitive parameter,
due to the cooling properties of the gas. Accretion of intergalactic matter,
which was not studied in this work, may prove a key ingredient to reproduce
the star formation history of spiral galaxies. It will be included in a future work.

Finally a simulation was run with a lighter dark matter halo, that is a less
stable initial configuration of the gas disk. As expected, the disc evolved
to form a bar and an $m=2$ spiral structure.

Our model takes a large range of physical processes into account. For some
of them, such as heating of the gas by the UV background, we use a very
simple algorithm. There is room for improvement at this level. But more
importantly, some of the physical processes, the star formation law in 
particular, are still not well understood. More simulation work is needed
to validate or reject the various criterion in use. Finally, there is the
important issue of modeling the cold gas dynamics. We used a simple sticky
particle scheme. Other choices (such as Andersen \& Burkert \cite{Andersen})
are worth investigating in association with the SPH warm gas dynamics. Another
possibility is to use SPH dynamics for the cold gas, but with an equation of
state different from the ideal gas. With the correct multi-phase 
description of the gas, it will be possible to study a wide range of phenomena
such a galaxy mergers or accretion, where cooling flows play an important
role.

\end{document}